\documentclass[aoas,MSNbibl,nameyear,seceqn,dvips]{arximspdf}
\usepackage{dcolumn}
\usepackage{graphicx}

\doi{10.1214/11-AOAS510}
\volume{6}
\issue{2}
\pubyear{2012}
\firstpage{521}
\lastpage{541}

\makeatletter

\let\sv@tabnotetext\tabnotetext
\let\sv@tabnotemark@fmt\tabnotemark@fmt
\long\def\legend#1{{\let\tabnote@indent\leavevmode\sv@tabnotetext[]{}{#1}}}

\newcolumntype{d}[1]{D{.}{.}{#1}}

\makeatother

\begin{document}
\begin{frontmatter}

\title{Statistical tests for the intersection of~independent
lists of genes: Sensitivity, FDR, and~type~I error control\thanksref{T1}}
\runtitle{List-intersection tests}

\thankstext{T1}{Supported by the Breast Cancer Research Foundation
(Barbara Parker, PI) and the National Institute of Health U54 HL108460.}

\begin{aug}
\author[A]{\fnms{Loki} \snm{Natarajan}},
\author[A]{\fnms{Minya} \snm{Pu}}
\and
\author[A]{\fnms{Karen} \snm{Messer}\corref{}\ead[label=e1]{kmesser@ucsd.edu}}
\runauthor{L. Natarajan, M. Pu and K. Messer}
\affiliation{University of California, San Diego}
\address[A]{Division of Biostatistics and Bioinformatics\\
Department of
Family and Preventive Medicine\\
Moores UCSD Cancer Center 0901\\
University of California, San Diego\\
La Jolla, California 92093\\
USA\\
\printead{e1}} 
\end{aug}

\received{\smonth{12} \syear{2010}}
\revised{\smonth{9} \syear{2011}}

%
\begin{abstract}
Public data repositories have enabled researchers to compare results
across multiple genomic studies in order to replicate findings. A
common approach is to first rank genes according to an hypothesis of
interest within each study. Then, lists of the top-ranked genes within
each study are compared across studies. Genes recaptured as highly
ranked (usually above some threshold) in multiple studies are
considered to be significant. However, this comparison strategy often
remains informal, in that type I error and false discovery rate (FDR)
are usually uncontrolled. In this paper, we formalize an inferential
strategy for this kind of list-intersection discovery test. We show
how to compute a $p$-value associated with a ``recaptured'' set of genes,
using a closed-form Poisson approximation to the distribution of the
size of the recaptured set. We investigate operating characteristics
of the test as a function of the total number of studies considered,
the rank threshold within each study, and the number of studies within
which a gene must be recaptured to be declared significant. We
investigate the trade off between FDR control and expected sensitivity
(the expected proportion of true-positive genes identified as
significant). We give practical guidance on how to design a
bioinformatic list-intersection study with maximal expected sensitivity
and prespecified control of type I error (at the set level) and false
discovery rate (at the gene level). We show how optimal choice of
parameters may depend on particular alternative hypothesis which might
hold. We illustrate our methods using prostate cancer gene-expression
datasets from the curated Oncomine database, and discuss the effects of
dependence between genes on the test.
\end{abstract}

%

\end{frontmatter}
\newpage
\section{Introduction}\label{intro}

Given several independent genomic data sets which address a similar
question, it is common to compare the lists of the top-ranked genes
from each study. Genes selected as highly ranked in multiple studies
may be considered validated or replicated. Curated databases of gene
lists are available which include tools for comparing lists and
intersecting lists of top-ranked genes across multiple similar studies
[\citet{Glez-Pena}, \citet{Culhane}]. The ``correspondence at the top'' concordance
statistic is an example of this approach \citet{Irizarry}. Perhaps the
most well known example is the study by Tomlins et al. of gene
expression in solid tumors, \citet{Tomlins} which compared the top 10
genes from each of 132 cancer studies in a publicly available
microarray data repository. Within each study, the genes were ranked
according to a statistic scoring potential ``fusion gene'' properties, as
such fusion genes are known to be important drivers of malignancy in
several hematologic cancers. Tomlins et al. targeted a
candidate gene list of 300 known cancer genes; any candidate gene which
ranked among the top 10 in \textit{two} or more of the studies was
considered to be a potential hit. Two significant genes were found, one
which ranked among the top 10 in two different studies and another in
five studies. For these two genes, a fusion product was subsequently
experimentally confirmed in prostate cancer and these remain the only
common fusion transcripts discovered in an epithelial tumor.

In this paper we show how to conduct such an intersection-of-lists
approach to assessing significance while controlling type I error (at
the set level) and false discovery rate (at the gene level). Given $N$
independent studies, there are two parameters which define a ``hit'': the
rank threshold,~$r$, above which a gene must lie in each study ($ r
=10$ in the Tomlins example), and the number of lists, $n$, among which
a gene must be ranked ($n=2$ in the Tomlins example) in order to be
declared significant. Our first goal is to define an exact $p$-value
which is easy to compute, when assessing the intersection of $n$ lists
of top-ranked genes, at rank $ r$ or above. This entails defining an
appropriate test statistic and corresponding hypothesis test, which we
call a~list-intersection discovery test, as this is an ``unsupervised''
or discovery approach. We apply these ideas to the related ``supervised''
case of an a~priori candidate gene list which is compared
against $N$ other independent studies, as in \citet{Tomlins}. Here the
aim is to validate the genes appearing in the researcher's a
priori list with a formal test of hypothesis. Following \citet
{Irizarry}, we call this a list-intersection concordance test. We then
develop practical guidelines for choices of~$r$ and~$n$ which maximize
the expected sensitivity at a given false discovery rate (FDR), that
is, which maximize the expected proportion of true-positive genes that
are declared to be significant at a given FDR. We give example
applications of both the discovery and concordance test using data from
the Tomlins study.

To state the discovery problem more precisely, consider data from $N$
independent (gene-expression) studies. Within each study, suppose $T$
genes are independent and are ranked according to a statistic, and
consider the list of the $ r$ top-ranked genes in each study. The set
of genes which lie in the intersection of $n$ or more of these lists,
$S_n( r)$, are those genes ``recaptured'' as significant at least $n$
times across $N$ independent studies. However, the degree of confidence
in this validation remains to be assessed. For example, considering
$N=6$ independent studies, each with $ 10\mbox{,}000$ possible genes, it may
be very likely that by chance alone at least 50 genes would be
recaptured among the top $r=200$ genes in $n = 2$ or more studies (as
we shall demonstrate, probability $0.84$), somewhat likely that 5 or
more genes would be so recaptured across $n=3$ studies (probability
$0.02$), and very unlikely that any one gene would be recaptured in
$n=5$ out of 6 studies just by chance (probability $0.0002$). In this
paper we show how to compute these probabilities (these examples are
computed in Section \ref{example1}), how to assess the statistical
significance of the recaptured set for given $ r$ and $n$, and how to
estimate the false discovery rate within the recaptured set. The test
statistic we use is $|S_n ( r)|$, the size of the intersection set; in
the above three examples $|S_n ( r)| \ge 50$, 5 and 1, respectively. In
making these computations we have assumed the genes behave
independently, which is no doubt not true in practice, and this is
addressed theoretically and in simulations in the last section of the paper.

The paper is organized as follows: in Section \ref{seccrstat} we derive
the distribution of the list-intersection test statistic under the null
hypothesis and show how to compute a $p$-value for a gene-set and how to
estimate the within-set false discovery rate (FDR). In Section \ref
{secConcordance} we derive the distribution of the list-intersection
concordance test statistic. In Section \ref{Tomex} we apply the test to
data in \citet{Tomlins}. In Section \ref{design}
we discuss how to control the type I error of the discovered set, and
how to control the false discovery rate of the genes within the
discovered set. We give a strategy for finding good choices of $ r$ and
$n$ (Section \ref{secbonferroni}). In Section \ref{secexample} we give
an example of how to mine a data repository for a ``statistically
significant'' discovered gene set while controlling the type I error at
the set level and the within-set FDR. Section \ref{conclusion}
addresses what happens if independence on genes does not hold, and then
gives conclusions and future directions. Simulation studies, code, and
additional proofs are described in the supplemental article
[\citet{Natarajan}].

\section{The list-intersection discovery test}\label{seccrstat}

The list-intersection test compares the top-ranked gene lists from
multiple studies in order to discover a common significant set of
genes. Suppose we consider $N$ studies, each of which investigate $T$
genes, and that the genes within a study are ranked according to a
prespecified scoring procedure which might be fold change, a between
group t-test, or might differ from study to study. Consider the list of
``top $ r$'' genes within each study, and consider the set of genes,
$S_n( r)$, which lie in $n$ or more of these top-ranked lists. (We will
often omit the dependence on $ r$ for convenience.) In this section we
find the expected count $\mathrm{E}_{H_0}|S_n|$ under the null
hypothesis of random ranking of the genes. We show that $|S_n|$ has an
approximately Poisson distribution under the assumptions the genes
within a study are independent, and that $T \gg r $ and $\mathrm{E}|S_n|
\ll r $ where both $T$ and $ r$ are large, and use this to compute a $p$-value.

\subsection{Null distribution of $|S_n(r)|$, estimated
FDR and $p$-values}\label{secnull}

For an arbitrary gene $g$, let the Bernoulli trial $B_g= 1$ if the gene
ranks among the top $ r$ genes in $n$ or more studies, with $B_g= 0$
otherwise. Then
%
\begin{equation}\label{distSn}
|S_n| = \sum_{g=1}^T B_g.
\end{equation}

Let $ P^0_n =P(B_g=1) $ denote the associated probability under the
null hypothesis. (Note that we have suppressed the dependence of
$P^0_n$ on $r$ for convenience.) While the Bernoulli trials $B_g$ are
not independent even under the assumption of independent genes (given
that one gene lies among the top $r$ genes, the next gene is less
likely to do so), they are identically distributed, and so it is
immediate that
%
\begin{equation}\label{ESn}
\mathrm{E}_{H_0}|S_n|=TP^0_n.
\end{equation}

To evaluate $P^0_n$ under the null hypothesis of random ranking among
$T$ genes in each study, index the studies from $i= 1 $ to $N $ and
again consider an arbitrary gene~$g$. Let $A_i$ be the Bernoulli trial
that counts a success if~$g$ ranks among the top $r$ genes in study
$i$, with $P(A_i =1)=p_i$. Under the null hypothesis, $p_i= r /T \equiv
p_0$, and the $A_i$ are independent. Let $X$ count the number of
successes for gene $g$. Then $X \sim\operatorname{Bin} (N,p_0) $, and the
probability~$P^0_n $ that gene $g$ is listed among the top $ r$ genes
in $n$ or more studies is given by
%
\begin{equation} \label{P0}
P^0_n = P(X \geq n),
\end{equation}
an easily computed binomial probability. Using (\ref{ESn}) and (\ref
{P0}), one may then estimate the within-set FDR by comparing the
expected number of discoveries under the null hypothesis to the total
number of discoveries made:
%
\begin{equation}\label{hatFDR}
\widehat{\mathrm{FDR}}=\mathrm{E}_{ H_0} [|S_n| ] /|S_n|.
\end{equation}
%

Under the null hypothesis of \textit{independent} random ranking of the
genes, we can derive the distribution of $|S_n|$. Note that for large
$T$ with $T \gg r$, selection of the $ r$ top-ranked genes within a
study has nearly the same distribution as random sampling with
replacement. If, in addition, \mbox{$\mathrm{E}|S_n|\ll r$}, then $B_g$ and
$B_h$ are approximately independent for any pair of genes~$g$ and~$h$.
In this case~$|S_n|$ will have an approximate Binomial distribution
with parameters~$T$ and~$P^0_n$. If, in addition, $T$ is large and
$P^0_n$ small, it follows that the distribution of $|S_n|$ is
approximately Poisson with mean $TP^0_n$. We consider the effects of
correlation between genes in Section~\ref{seccorr}.

\subsection{Example computations using the list-intersection statistic}
\label{example1}
Here we show how to use (\ref{ESn}) and (\ref{P0}) to compute the
expected number of genes recaptured just by chance, as well as the
$p$-value of the size of the recaptured set and the estimated FDR for
genes within the set. These quantities depend on the total number of
studies considered, $N$, the depth of the top-ranked list, $r$, and the
number of lists intersected, $n$. Throughout we let $T=10\mbox{,}000$. We also
investigate the quality of the Binomial and Poisson approximations for
these examples:

\begin{longlist}
\item As in the \hyperref[intro]{Introduction}, consider that we have ranked
$T=10\mbox{,}000$ genes, and that the top $r=200$ genes are the top-ranked
set. Then, under the null hypothesis of independent random ranking,
$p_0 = r/T = 0.02$. From~(\ref{P0}), given $N= 6$ studies to compare,
the probability of seeing a given gene in the top 200 from $n=2$ or
more studies is $P^0_2=0.0057$. It follows from~(\ref{ESn}) that
$|S_2(200)|$, the number of genes captured in 2 or more studies, is
then approximately Poisson with mean $T P^0_2= 57$.

To evaluate the accuracy of $p$-values computed from this Poisson
approximation, note that two of the three key assumptions, that $T \gg
r$ and $P^{0}_n$ be small, are met. However, $E |S_2| /r \approx0.30$,
so that $E |S_2| $ is not particularly small compared to~$r$. Suppose
the observed value of $ |S_2(200)|=50$. Then, from simulation under the
null hypothesis of random ranking, $P(|S_2(200)| \geq50) = 0.86$,
compared to the corresponding Poisson $p$-value of $0.84$, yielding
a~relative error of $2.3\%$. Further examples show the simulated 95th
percentile of the null distribution is 68, while the Poisson
approximation gives~70 (relative error, $2.9\%$). At 1\% significance
level, the relative error is 2.7\% (simulated value 73; Poisson
approximation~75). Thus, the Poisson approximation appears to work well
in this case.

\item Continuing the example, if we require the genes to be
recaptured in three or more studies (so that $n=3$ rather than 2), the
mean number of genes captured under the null is only 1.53. As in the
\hyperref[intro]{Introduction}, suppose the observed value of $ |S_3(200)|=5$. Under the
null hypothesis of random ranking the probability that 5 or more genes
would be in the intersection list is $P(|S_3(200)| \geq5 )=0.02$,
where $|S_3(200)|$ is Poisson with mean 1.53. Thus, we would have seen
a statistically significant event with a $p$-value of 0.02. The estimated
within-set FDR would be $1.53 /5$, or 31\%. Note that from simulation,
$P(|S_3(200)| \geq5 ) =0.018$, again demonstrating the adequacy of the
Poisson approximation.

\item Now suppose only $N=4$ rather than $6$ total studies are
considered, and $n=2$ as before. Then the expected number of genes
captured by chance falls by half, to a mean of 23 genes.

Again the relative error of the 95th and 99th percentiles of
$|S_2(200)|$ from the Poisson approximation is $3\%$ (the simulated
95th percentile of the null distribution is 31, while the Poisson
approximation gives 32; the simulated 99th percentile is~34, while the
Poisson approximation gives 35).

\item When $N=6$ but the depth of the list is halved so that only
the top $r=100$ genes are considered, the mean number of genes captured
by chance falls by $3/4$, from 57 to 15. The relative approximation error
of the Poisson distribution is 0\% for the 95th percentile, and 4\% for
the 99th percentile.
\end{longlist}

These examples show how to use the Poisson approximation to the
distribution of $S_n(r)$ to calculate $p$-values and FDRs. Over the range
of parameters considered here, the Poisson approximation appears to be
very good. Additional simulations are reported in Section \ref
{secexample} and Section 1.1 of the supplemental article
[\citet{Natarajan}].

\section{The list-intersection concordance test}
\label{secConcordance}

The concordance test evaluates whether an a priori candidate
list of $m$ genes, say, from the researcher's new study, is
significantly reproduced among the top $r$ genes in $N$ independent
ranked lists of genes, say, from other experiments or from the
literature. Suppose each study investigates $T$ genes and consider the
set of genes, $C^m_n( r)$, from the a priori candidate list
which also lie in $n$ or more of these top-ranked lists. As before, we
show that $|C^m_n|$ has an approximately Poisson distribution under the
null hypothesis of independent random ranking of the genes, however,
with a different mean, under the assumptions that $T\gg r $ and $\mathrm
{E}|C^m_n| \ll r$ and both $T$ and $ r$ are large.

\subsection{Null distribution of $|C^m_n(r)|$}
\label{secnull2}

Again, index the studies from $i= 1 $ to~$N$. Consider an arbitrary
gene $g $ drawn from the a priori list of $m$ genes of
interest, and, as before, for study $i$ let $A_i$ be the event that
gene $g$ is listed among the top $ r$ genes. Under the null hypothesis
of random ranking among $T$ genes in each study, $p_i= r /T \equiv
p_0$, as before.
As in Section \ref{secnull}, equation (\ref{P0}) gives $P^0_n$, the
probability under the null hypothesis that $n$ or more of the events
$A_1,\ldots, A_N$ occur simultaneously. Now consider the~$m$ genes on
the a priori list, and let $B_g= 1$ if the $g$th gene ranks
among the top $ r$ genes in $n$ or more studies, with $B_g= 0$
otherwise. Under the null hypothesis $P(B_g=1) =P^0_n$, and as $|C^m_n|
= \sum_{g=1}^m B_g$, it is immediate that
%
\begin{equation}\label{Cmn}
\mathrm{E}|C^m_n|=mP^0_n
\end{equation}
under the null. Further, for large $T$ with $T \gg r$, selection of the
$ r$ top-ranked genes within a study has nearly the same distribution
as random sampling with replacement. If in addition $\mathrm
{E}|C^m_n|\ll r$, then $B_g$ and $B_h$ are approximately independent for
any pair of genes $g$ and $h$. In this case $|C^m_n|$ will have an
approximate Binomial distribution with parameters $m$ and $P^0_n$,
which in turn is approximately Poisson with mean $mP^0_n$ for $m$ large
and $P^0_n$ small.

\subsection{Example test using data from Tomlins et al}
\label{Tomex}

We apply these computations to the data from \citet{Tomlins}.
They considered the Cancer Gene Census [\citet{Futreal}] published list of
300 genes known to be involved in cancer, and compared this candidate
gene list across 132 studies from the Oncomine [\citet{Onco}] repository
of microarray data. Within each study, they ranked all genes according
to a score characteristic of a fusion gene. They then looked for the
occurrence of any candidate cancer genes among the 10 top-ranked genes
in each study, and for each cancer gene, reported how many times it was
``captured'' in a top-10 list. To define parameters, each microarray
platform interrogated about 10,000 expressed genes. Thus, we have
$T=10\mbox{,}000$ genes across $N=132$ studies, with the top $r=10$ genes
considered from each study. The length of the a priori list
is $m=300$.

%
\begin{table}
\caption{{Example: expected and observed number of recaptured
candidate genes, $p$-values and estimated FDR within the recaptured
set.} Data from Tomlins et~al. (\protect\citeyear{Tomlins}); $T=10\mbox{,}000$
genes across $N=132$ studies, with the top $r=10$ genes considered from
each study. The length of the a priori candidate gene list is $m=300$.
The recapture rate $n$ varies from 2 to 5; $n=2$ was the choice used in
Tomlins et~al. (\protect\citeyear{Tomlins})}
\label{Tom}
\begin{tabular*}{\tablewidth}{@{\extracolsep{\fill}}l cd{1.3}cc@{}}
\hline
& $\bolds{n=2}$ & \multicolumn{1}{c}{$\bolds{n=3}$} & \multicolumn{1}{c}{$\bolds{n=4}$}
& \multicolumn{1}{c@{}}{$\bolds{n=5}$} \\
\hline
$E|C_n| $ under null& 2.5\hphantom{0}& 0.11& $<$0.01 & $<$0.01 \\
[2pt]
Observed $C_n$ & \{ERBB2, ERG, ETV1, IRTA1\} & \multicolumn{1}{c}{ERG} & ERG &ERG \\
[2pt]
Observed $|C_n|$ & 4\hphantom{.00}& 1 & 1 & 1 \\
[2pt]
$p$-value & 0.25 & 0.006 & $7\times10^{-6}$ & $5\times10^{-9}$ \\
[2pt]
Estimated FDR & 0.63 & 0.11 & $<$0.01 & $<$0.01 \\
\hline
\end{tabular*}
\end{table}

We applied (\ref{P0}) and (\ref{Cmn}) to find the expected number of
cancer genes which appear in the intersection of $n$ multiple lists
under the null hypothesis of independent random ranking, for $n$
ranging from 2 (the case considered by Tomlins et al.) to 5. These
results are given in Table \ref{Tom}, in the row labeled $E|C_n|$.
We give the set of actual genes found by Tomlins et al. in the
intersection of $n$ or more lists (the observed set $C_n$), taken from
their supplementary Table S1. We also record the count of cancer genes
recaptured $n$ or more times (the observed count $|C_n| $). We then
compute the $p$-value for each value of $n$, computed as the probability
that a Poisson variate with the given mean would lie above the observed
value of $|C_n| $, and the estimated FDR within each recaptured set.
Notably, the observed set of cancer genes which is in the intersection
of 2 or more lists, the set considered by Tomlins et al., has
a $p$-value of 0.25, indicating it is plausible that this many genes
would reappear just by chance. Four genes were ``discovered'' in this
recaptured set, while the expected number recaptured under the null is
2.5 for an estimated FDR of $2.5/4$ or 63\%. However, the $p$-values
attached to the single gene ERG, which reappears in 5 studies, is
highly significant. Both ERG and the related ETV1 were subsequently
validated as fusion genes.

This example illustrates how to compute the $p$-value for the size of an
observed set of concordant genes. However, notice that multiple
$p$-values are presented in Table \ref{Tom}, corresponding to multiple
choices of $r$ and $n$. Unless we specify $r$ and $n$ in advance, we
are open to charges of data snooping, that is, of tailoring the choice
of $r$ and $n$ to the results they yield in a given data set, rendering
the nominal $p$-values invalid. Thus, this example also highlights the
need for a strategy for choosing $r$ and $n$, and, importantly, the
need to specify the choice of $r$ and $n$ before the analysis is
carried out. We discuss these issues in the remainder of the paper.

\section{Control of type I error and within-set FDR} \label{design}

For a given prespecified choice of $r$ and $n$, the list-intersection
test will declare a gene set to be significant only if $|S_n(r)|$ [or
$|C_n^m(r)|$] has a $p$-value below the stated significance level $\alpha
$. This procedure will strictly control the type I error rate; that is,
under the null model, the probability will be at least $1-\alpha$ that
no gene set will be declared to be significant. Given a statistically
significant gene set, it remains to investigate the FDR within set, and
the expected proportion of true positive genes that are captured (the
expected sensitivity).

Importantly, as noted above, control of type I error requires both $n$
and $r$ to be specified in advance. For example, there may be several
sets $S_n(r)$ with $p$-values falling below any given significance level,
and post hoc selection of one or more of these sets without
correction for multiple testing would of course leave both the type I
error and the set-level FDR uncontrolled. In addition, failure to
prespecify $r$ and $n$ is likely to lead to data snooping, in which the
chosen $r$ and $n$ are consciously or unconsciously tailored to yield
the most ``interesting'' set of selected genes. Thus, it remains to
consider how to make good a priori choices of $r$ and $n$. In
this section, we give an example which illustrates how good choices of
$r$ and $n$ may depend on which particular alternative hypothesis
holds, and then propose a general design strategy. We leave as future
work discussion of the more computationally and mathematically involved
data-driven strategies to control the FDR.

\subsection{Example choices of $r$ and $n$:
Expected sensitivity and false discovery rate}\label{secchoosingXandj}

Different choices of the threshold $ r$ and the recapture rate $n$
will trade off between an increased false discovery rate within the set
$S_n(r)$ and increased power to capture any truly positive genes. For
example, for fixed $n$, as $r$ increases and more genes are included in
the set of ``top-$r$'' genes, any truly significant genes (``true
positives'') will be more likely to be selected within each study and
thus more likely to land in the intersection set $S_n(r)$. However, at
the same time more null genes will be captured, thereby potentially
increasing the FDR within $S_n(r)$, and possibly reducing power to call
$S_n(r)$ a statistically significant set. Good choices for~$r$ and~$n$
will evidently depend on how many truly positive genes exist, as well
as the effect size for each, as the latter determines the probability
that a given positive gene rises to the top of the list.

%
\begin{table}
\caption{Expected sensitivity (ESns) and FDR under different
alternative hypotheses and choices of $ r$ (within-study significance
threshold), and $n$ (recapture rate)}
\label{fdr}
\begin{tabular*}{\tablewidth}{@{\extracolsep{\fill}}lcd{3.3}d{2.3}d{2.3}d{2.3}d{2.3}@{}}
\hline
\textbf{Recapture-rate ($\bolds{n}$)} & & \multicolumn{1}{c}{$\bolds{r =500}$}
& \multicolumn{1}{c}{$\bolds{r = 100}$} & \multicolumn{1}{c}{$\bolds{r =50}$}
& \multicolumn{1}{c}{$\bolds{r =25}$} & \multicolumn{1}{c@{}}{$\bolds{r = 10}$} \\
\hline
\multicolumn{7}{@{}c@{}}{{Alternative hypothesis I: \# of true
positive genes$ {}={} $25, each with 3-$\sigma$ upregulation}}\\
\multicolumn{7}{@{}c@{}}{{as
compared to null genes}} \\[4pt]
2 & EFP& 128.43 & 3.99 & 0.71& 0.10& 0.003 \\
& ETP& 24.93 & 23.36 & 21.02 &16.91 & 9.05 \\
& ESns &0.99 & 0.93 & 0.84 & 0.68 & 0.36 \\
& FDR& 0.84 & 0.15 & 0.03 & 0.006 & \mbox{$<$}0.001\\
& & & & & & \\
3 & EFP& 4.21 & 0.02 & 0.002 &  \mbox{$<$}0.001 &  \mbox{$<$}0.001 \\
& ETP& 23.90 & 17.42 & 12.65 & 7.54 & 2.24 \\
& ESns &0.96 &0.70 &0.51& 0.30& 0.09 \\
& FDR& 0.15 & 0.001 &  \mbox{$<$}0.001 & \mbox{$<$}0.001 & \mbox{$<$}0.001 \\
& & & & & & \\
4 & EFP & 0.05 &  \mbox{$<$}0.001 &  \mbox{$<$}0.001 &  \mbox{$<$}0.001 &  \mbox{$<$}0.001 \\
& ETP & 17.06 & 6.94 & 3.64 & 1.47 & 0.22 \\
& ESns & 0.68 &0.28 &0.15 &0.059& 0.009 \\
& FDR& 0.003 &  \mbox{$<$}0.001 &  \mbox{$<$}0.001 &  \mbox{$<$}0.001 &  \mbox{$<$}0.001 \\
[4pt]
\multicolumn{7}{@{}c@{}}{{Alternative hypothesis II: \# of true
positive genes$ {}={} $2, each with 4-$\sigma$ upregulation}}\\
\multicolumn{7}{@{}c@{}}{{as
compared to null genes}} \\
[4pt]
2 & EFP& 139.14 & 5.70 & 1.38 & 0.32 & 0.04 \\
& ETP& 2.00 & 2.00 & 2.00 & 1.99 & 1.95 \\
& ESns &1.00& 1.00 &1.00 &1.00 &0.98 \\
& FDR& 0.99 & 0.74 & 0.41 & 0.14 & 0.02 \\
& & & & & & \\
3 & EFP& 4.76 & 0.04 & 0.005 & \mbox{$<$}0.001 & \mbox{$<$}0.001 \\
& ETP& 2.00 & 1.97 & 1.93 & 1.85 & 1.65 \\
& ESns &1.00& 0.99 &0.97 &0.93 &0.83 \\
& FDR& 0.70 & 0.02 & 0.002 & \mbox{$<$}0.001 & \mbox{$<$}0.001\\
& & & & & & \\
4 & EFP& 0.06 & \mbox{$<$}0.001 & \mbox{$<$}0.001 & \mbox{$<$}0.001 & \mbox{$<$}0.001\\
& ETP& 1.93 & 1.64 & 1.44 & 1.19 & 0.84 \\
& ESns &0.97& 0.82 &0.72 &0.60 &0.42 \\
& FDR& 0.03 & \mbox{$<$}0.001 & \mbox{$<$}0.001 & \mbox{$<$}0.001 & \mbox{$<$}0.001\\
\hline
\end{tabular*}
\legend{Note:
$N = 4$ independent studies; $T = 10\mbox{,}000$ genes measured in each study;
EFP${} = {}$expected \# of false positives; ETP${} = {}$expected \# of true
positives; ESns${} = {}$ETP$/$\# true positives; FDR${} =
{}$EFP$/$(EFP${}+{}$ETP).}
\end{table}

To illustrate these trade-offs, in Table \ref{fdr} we compute the expected
number of true discoveries and false discoveries for several choices of
$r$ and $n$, under two simple alternative hypothesis scenarios. We
considered $N=4$ independent studies, each investigating $T=10\mbox{,}000$
genes. We assume that the statistic used to rank the genes has an
approximately normal null distribution, such as a two sample
$t$-statistic or a maximum likelihood statistic. We assume a total of
$tp$ genes are true positives, and for each such gene, the statistic is
assumed to be normally distributed with mean $\mu$ and standard
deviation 1. We constructed two scenarios: alternative I had 25 true
positive genes, each upregulated by 3 standard deviations as compared
to the null genes, with the remaining $T-tp$ constituting the null
genes. In alternative~II, we considered $tp=2$ true positive genes,
each with expression levels upregulated by 4 standard deviations as
compared to the null genes. Thus, alternative I has multiple
significant genes, each with moderate effect-sizes, and alternative II
has a few true hits with large effect-sizes. Under each of these
illustrative alternative hypotheses, we computed the expected number of
null and significant genes recaptured by the list-intersection
statistic. The mathematical argument for these expectations uses an
argument from \citet{Feller} and is given in the supplement
(Section 2) [\citet{Natarajan}]. For our two chosen scenarios, and for
given $r$ and $n$, Table \ref{fdr} displays the FDR within the
intersection gene set as well as the expected sensitivity (the expected
proportion of true-positive genes that are captured). We considered
recapture rates $n$ from $2$ to $4$, and within-study thresholds $r$
from $500$ to $10$.

For alternative hypothesis I (25 true-positive genes, each with $3\sigma
$ upregulation), when $n=2$, a high expected sensitivity can be
achieved by choosing~$r$ to be large. For example, $r= 500$ has an
expected capture rate of 24.93 true positives out of 25 total, for an
expected sensitivity of 99.7\%. However, this is at the cost of an FDR
of over 80\%, as the expected number of false positives is over $128$
with a total expected set size of 153.36. Hence, the pair $ r=500$, $n=2$
does not appear to be a good choice here. Lowering $r$ from 500 to 100
reduces the expected number of false positives to $4$ while maintaining
the expected number of true positives captured at about 23 out of 25
(92\% expected sensitivity); thus, $(r=100, n=2)$ appears to be a~reasonable choice. Lowering $ r$ further achieves a lower FDR, but at
the cost of lower expected sensitivity: with $n=2$ as $ r $ decreases
from 50 to 10, the expected sensitivity decreases from 84\% to 36\%.
A better trade-off would be to require a larger recapture rate with $n
= 3$ while maintaining $ r =500$, as this combination maintains a
sensitivity of 95.6\% ($\mbox{ETP}= 23.9$ out of 25) while reducing false
discoveries ($\mbox{EFP} = 4.2$ and $\mbox{FDR} = 15$\%). Requiring a~recapture rate of 4
out of 4 studies is too stringent for the scenario considered here.
Thus, either $( r=100, n=2)$ or $( r=500, n=3)$ appear to be good
choices for alternative hypothesis I; both have expected sensitivity
over 90\% and FDR under 15\%. Note that it may not always be possible
to achieve high sensitivity and low FDR; in this case, as is evident in
Figure \ref{figdesign} below, the number of studies $N$ must be increased.

%
\begin{figure}

\includegraphics{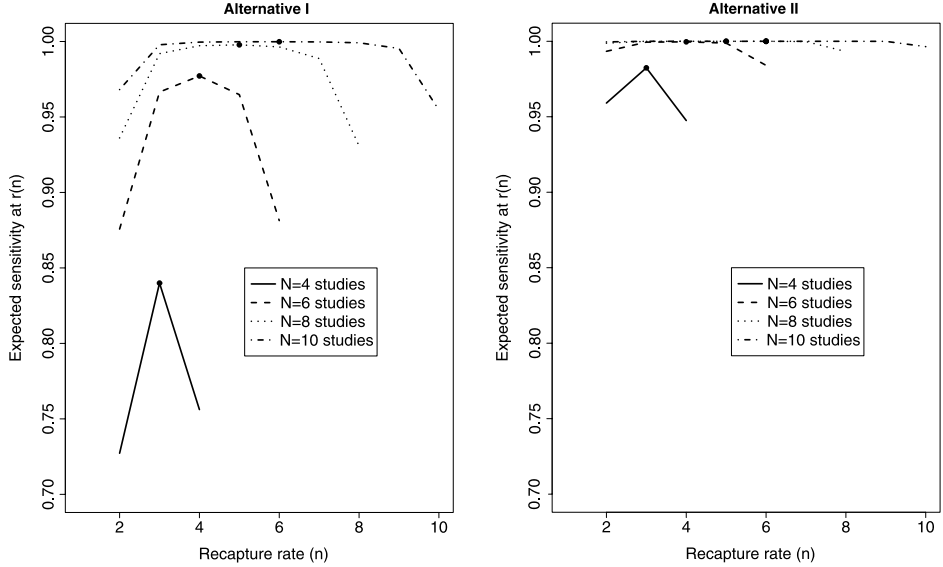}

\caption{{Expected sensitivity at $r(n)$ versus
recapture rate $n$, given $N$ studies in total and FDR${}\leq{}$0.01,
under two alternative scenarios.} For each recapture rate $n$, the
expected sensitivity at the optimal threshold $r(n)$ is plotted. The
filled circle represents the maximum expected sensitivity achievable
for a given $N$. As expected, sensitivity increases as the total number
of studies $N$ increases. The recapture rate ($n$) and threshold
[$r(n)$] which maximize sensitivity represent the optimal design
choices which control FDR at the prescribed rate. Here \# of
features$/\mbox{genes}=T=10$,000, and $\mbox{FDR}\leq0.01$;
the value of $r(n)$ is from Table~\protect\ref{tabdesign}. Left:
alternative \textup{I}: 25 true-positive genes each with effect-$\mbox{size} =
3\sigma$. Optimal design
for \textup{(i)} $N=4$ at $n = 3$, $r= 195$; \textup{(ii)} $N=6$ at
$n = 4$, $r= 384$; \textup{(iii)}~$N=8$ at $n = 5$, $r= 575$; \textup{(iv)} $N=10$ at $n = 6$, $r= 755$.
Right: alternative \textup{II}: 2 true-positive genes each with effect-size~$4\sigma$.
Optimal design for \textup{(i)} $N=4$ at $n = 3$, $r= 81$; \textup{(ii)}
$N=6$ at $n = 4$, $r= 194$; \textup{(iii)} $N=8$ at $n = 5$, $r= 332$;
\textup{(iv)} $N=10$ at $n = 6$, $r=475$.} \label{figdesign}
\end{figure}

For alternative hypothesis II (2 true positive genes, each upregulated
by~$4 \sigma$), for a recapture rate of $n=2$, thresholds of $ r = 500,
100$ or $50$ all have expected sensitivity of 100\%, but also have FDR
over 40\%. (However, note that the \textit{number} of false discoveries
may not be prohibitive.) A cutoff of $ r = 25$ or $r=10$ gives better
control of the FDR, while maintaining a high expected sensitivity. When
$n=3$, stringent thresholds such as $ r =25$ or $10$ result in capture
of fewer true positives, whereas setting $ r = 100$ appears to be a
good trade-off. Again, requiring a recapture rate of $n=4$ reduces the
expected sensitivity, so the reasonable pairs among those considered
appear to be either $( r=100, n=3)$ or $( r=10, n=2)$.

We have illustrated how equations (\ref{P0}) and Supplement (2.1)
[\citet{Natarajan}] can be used to calculate the expected number of true and
false positives, and FDR and expected sensitivity for various
postulated hypotheses. In the next section we examine how these methods
might be applied when designing a bioinformatic search to test
a priori hypotheses of interest.

\subsection{\texorpdfstring{Choosing $r$ and $n$ to maximize
sensitivity, with $\mbox{FDR}\leq q$}{Choosing r and n to maximize
sensitivity, with FDR <= q}}\label{secbonferroni}

The example in Section \ref{secchoosingXandj} illustrates that the
best choice of threshold $ r$ and recapture rate $n$ will depend on the
number of true positive genes, as well as the effect size for these
genes. These considerations suggest how to design a~list-intersection
test: given an acceptable FDR $q$, find $ r$ and $n$ that maximize the
expected sensitivity while maintaining the gene-wise $\mbox{FDR} \leq q$. This
can be computed for a prespecified alternative hypothesis which
postulates $tp$ true positive genes and corresponding effect-sizes as
outlined below:
\begin{enumerate}[(2)]
\item[(1)] Set an acceptable $\mbox{FDR} = q$.
\item[(2)] For each possible recapture rate $n = 1,2, \ldots, N$, find the
maximum threshold $r(n)$ which still maintains $\mbox{FDR}<q$:
\begin{enumerate}[(a)]
\item[(a)] For each $r = 1, \ldots, T$:
\begin{enumerate}[(iii)]
\item[(i)] compute the expected number of recaptured false positive genes
$\mbox{EFP}(n,r)=(T-tp)P_n^0 (r)$ [see Supplement equation (2.1)].
\item[(ii)] Given $tp$ true positive genes and their effect-sizes,
calculate
$\mbox{ETP} (n,r)$, the expected number of recaptured true positive genes.
This can be obtained using Supplement equation (2.1) as $\mbox{ETP}(n,r) = \sum
_{a=1}^{tp}
P_n^a(r)$.
\item[(iii)]
Calculate FDR$(n,r)$ as $ \frac{\mathrm{EFP}(n,r)}{\mathrm{EFP}(n,r) + \mathrm{ETP}(n,r)}$.
\end{enumerate}
\item[(b)] Let $r(n) = \max_{r=1}^{T} \{r| \mathrm{FDR}(n,r) < q
\}$.
\end{enumerate}
\item[(3)] For each pair $(n, r(n))$, calculate its expected sensitivity,
$\mbox{ETP}(n,r(n))/tp$.
\item[(4)] Choose the optimal pair, $(n,r(n))$, as the pair for which this
expected sensitivity is maximized.
\end{enumerate}

%
\begin{table}
\caption{The maximum threshold $r(n)$ that satisfies $\mbox{FDR} \leq
0.01$, for a given number of studies $N$ and recapture rate $n$. Two
alternative scenarios, $T= 10\mbox{,}000$ genes. Expected sensitivity of the test
at the maximum threshold is plotted in Figure \protect\ref{figdesign},
as a
function of $n$}
\label{tabdesign}
\begin{tabular*}{\tablewidth}{@{\extracolsep{\fill}}lcd{4.0}d{3.0}c@{}}
\hline
& \multicolumn{4}{c@{}}{\textbf{Total \# of studies ($\bolds{N}$)}} \\[-4pt]
& \multicolumn{4}{c@{}}{\hrulefill}\\
\textbf{Recapture rate ($\bolds{n}$)} & \multicolumn{1}{c}{\textbf{4}}
& \multicolumn{1}{c}{\textbf{6}} & \multicolumn{1}{c}{\textbf{8}}
& \multicolumn{1}{c@{}}{\textbf{10}} \\ \hline
\multicolumn{5}{@{}c@{}}{{Alternative hypothesis I:}} \\
\multicolumn{5}{@{}c@{}}{{25 significant genes, each upregulated by
3-$\sigma$}} \\[4pt]
2& \hphantom{0}30 & 24 & 20 & \hphantom{0}18 \\
3& 195 & 126 & 95 & \hphantom{0}77 \\
4& 683& 384 & 270 & 210 \\
5& -- & 869 & 575 & 434 \\
6& -- & 1698 & 1034 & 755 \\
[4pt]
\multicolumn{5}{@{}c@{}}{{Alternative hypothesis II:}}\\
\multicolumn{5}{@{}c@{}}{{2 significant genes, each upregulated by
4-$\sigma$}} \\
[4pt]
2& \hphantom{00}7 & 5 & 4 & \hphantom{00}3 \\
3& \hphantom{0}81 & 48 & 34 & \hphantom{0}27 \\
4& 373 & 194 &133 & 102 \\
5& -- & 513 &332 & 247 \\
6& -- &1123 &659 & 475 \\
\hline
\end{tabular*}
\end{table}

To illustrate this strategy, we again examined the two alternative
scenarios discussed in Section \ref{secchoosingXandj}. We set $T=
10$,000, as before, and let $N$ vary from 4 to 10 studies. We bounded
the FDR by $q=0.01$. Table \ref{tabdesign} lists the maximum threshold
$r (n)$ which satisfies the FDR bound, as obtained from step 2 of the
above algorithm, for each possible recapture rate $n$. Note that $r(n)$
increases rapidly with increasing $n$. For example, under alternative
hypothesis I, with $N=4$ studies and $n=2$, the maximum threshold which
maintains the FDR cutoff is $r(2)=30$, whereas if we consider
intersections across all $4$ studies (i.e., $n=4$), the maximum
threshold is, as expected, larger at 683, since null genes will be less
likely to be recaptured across all studies. Note that $r(n)$ decreases
as the number of studies $N$ increases since the chance of a false
positive increases with the total number of studies and, hence, the
size of the recaptured list would need to be smaller to satisfy the
prespecified FDR.

For each given number of total studies $N$, Figure \ref{figdesign}
plots the expected sensitivity, $\mbox{ETP}/tp$, against $n$ for the optimal $
r(n) $ from Table \ref{tabdesign}. Given $N$ studies in total, the pair
$(n,r(n))$ that maximizes sensitivity would be the optimal a
priori design choice for the study. For instance, under alternative
hypothesis I and an FDR cutoff of $0.01$, with $N=4$ total studies, the
maximal expected sensitivity of $\sim$84\% is achieved at recapture
rate $n = 3$, which from Table \ref{tabdesign} is achieved at threshold
$r(3)=195$. The other two scenarios corresponding to $(n,r(n)) =
(2,30)$ or $(4,683)$ achieve an expected sensitivity of less than
$80\%$. Hence, for alternative I and 4 total studies [$n=3,r(3)=195$] is
the optimal design choice. Note that under a given alternative, as the
total number of studies $N$ increases, the best choice of recapture
rate $n$ increases, as does the expected proportion of true positive
genes recaptured [the expected sensitivity at the optimal choice of
$(n,r(n)]$.

The calculations in Table \ref{tabdesign} and Figure \ref{figdesign}
illustrate how a good choice of $ r$ and $n$ involves maintaining
control of the FDR while maximizing the chance of capturing true
positive genes. The best choice of the pair $(n, r(n))$ of course
depends on whether one expects many significant genes with
small-moderate effect sizes similar to alternative I, or few
differentially expressed genes at large effect sizes, similar to
alternative II.
For a given alternative hypothesis, our design strategy chooses the
optimal combination of~$r$ and~$n$ which maximizes the expected
sensitivity, while controlling the FDR at the desired level. Note that,
if the expected sensitivity at the optimal pair $(n, r(n))$ is not
satisfactory, then either the number of studies considered $N$ must be
increased or the desired FDR level must be relaxed.

If multiple alternatives are proposed with no clear ``winner,'' the
above procedure can be used to choose the optimal design for several
proposed alternatives. Then a Bonferroni correction could be applied,
and the gene-sets that pass a Bonferroni corrected significance level
would be candidates for further research. Specifically, for a given
alternative and optimal design choice $(n,r(n))$, a $p$-value can be
calculated for each $|S_n(r(n))|$ the test statistic of the observed
data. This $p$-value might be computed using the approximate Binomial or
Poisson distributions (Section \ref{seccrstat}) or via simulation. Then
for $m$ possible alternatives, and a significance level $\alpha$, the
gene-sets for which the corresponding $p$-values are less than $\alpha/m$
are considered ``significant.'' This procedure strictly controls the
type I error rate on the selected significant sets. Thus, under the
null model, the probability is $\alpha$ or less of declaring a set of
genes to be significant.

\section{Example: Mining the Oncomine database for
candidate fusion genes in prostate cancer}\label{secexample}

To illustrate our methods, we carried out an example list-intersection
discovery study using the publicly available Oncomine database
[\citet{Onco}], as in the original Tomlins study
[\citet{Tomlins}]. We identified 4 suitable microarray gene
expression prostate cancer studies [\citet
{Dhan}, \citet{Lap}, \citet{Tiwari}, \citet{Tomlins2}];
our selection criterion was that all use a
similar cDNA microarray platform, and, as is common in such studies
[\citet{Tomlins}], we assumed that $\sim$10,000 genes would be
expressed. Thus, we have $N=4$ and $T= 10\mbox{,}000$. Note that here, unlike
in \citet{Tomlins}, we are conducting a genome-wide discovery
test, rather than a~concordance test based on a list of candidate
cancer genes. A second major difference is that Tomlins et al.
considered $N=132$ studies across all cancer types and all platforms. Of
course, Tomlins et al. used an ad hoc strategy to select interesting
sets of genes for validation, rather than the statistically motivated
use of $p$-values we are illustrating here. In this setting we know
there to be least two true positive genes (the fusion genes identified
in Tomlins et al.), and we are interested to see whether our a priori
search strategy will find them.

We set the significance level to $\alpha= 0.05$. We next will
prespecify the pair $(n, r(n))$, in order to avoid data snooping. Thus,
under the null hypothesis there would be only 5\% probability that the
study will declare any gene set to be significant (see example in Section \ref
{Tomex} and Section \ref{design}). As discussed in Section \ref
{secbonferroni}, good choice of $r$ and $n$ depends on the particular
alternative hypothesis postulated. Because in this somewhat artificial
setting we know that fusion genes are rare and that at least two exist,
alternative hypotheses~II (see Table \ref{fdr}) with $2$ significant
genes, each with an effect-size of $4$, is a~reasonable choice for our
study design. As in Table \ref{fdr}, we chose a stringent FDR cutoff of
$0.01$, with the rationale that then all discoveries within
a~significant set are likely to be true. Under these conditions, the
optimal design choice is $(n=3, r=81)$, so that the set of genes
$S_3(81)$ will be tested for statistical significance (Table \ref
{tabdesign} and Figure \ref{figdesign}).

%
\begin{table}
\caption{{List-intersection discovery of fusion-gene candidates
across $N=4$ independent prostate cancer studies.}
The test statistic $|S_n( r)| $ is the observed number of genes that
are ranked among the top $ r $ genes in at least $n$ studies;
$p$-value represents the probability of observing $|S_n( r)|$ or more
genes under the null hypothesis of independent random ranking. The
entry in bold corresponds to our a priori choice of $n=3$,
$r=81$; as the $p$-value for this entry is less than $ 0.05$, the
corresponding gene set is declared significant}
\label{ex}
\begin{tabular*}{\tablewidth}{@{\extracolsep{\fill}}lcd{2.2}d{2.2}d{1.3}d{1.4}d{2.4}@{}}
\hline
\textbf{\# of studies ($\bolds{n}$)} & & \multicolumn{1}{c}{$\bolds{r=100}$}
& \multicolumn{1}{c}{$\bolds{r=81}$} &
\multicolumn{1}{c}{$\bolds{r =50}$} & \multicolumn{1}{c}{$\bolds{r =25}$}
& \multicolumn{1}{c@{}}{$\bolds{r=10}$} \\
\hline
& $|S_2( r )|$ & 10& 6 & 4& 4& 2 \\
2 & $p$-value & 0.08 &0.20 & 0.06 & 0.0006 & 0.002 \\
& est.FDR\tabnoteref{ta} & 0.59 & 0.65 & 0.37 & 0.09 & 0.03 \\
[4pt]
& $|S_3( r )|$ & 1 & \multicolumn{1}{c}{\hspace*{3pt}\textbf{1}\hphantom{00}}&1 &1 &1 \\
3 & $p$-value & 0.04 & \multicolumn{1}{c}{\hspace*{5pt}\textbf{0.02}}& 0.005& 0.0006 &\mbox{$<$}0.0001 \\
& est.FDR\tabnoteref{ta} & 0.04 & 0.02 & 0.005 & 0.0006 & \mbox{$<$}0.0001 \\
[4pt]
& $|S_4( r )|$ & 0& 0&0& 0& 0 \\
4 & $p$-value &1 &1 &1 &1 & 1\\
\hline
\end{tabular*}
\tabnotetext[\mbox{$\ast$}]{ta}{est.FDR${} = {}$estimated
$\mbox{FDR} = E|S_n(r)|/|S_n(r)|$.}
\end{table}

Next, within each of the 4 identified prostate cancer studies, we
ranked the genes according to the ``cancer outlier profile analysis''
(COPA) procedure implemented in the Oncomine website [\citet{Onco}]. This
statistic measures ``fusion-like'' properties, and was used by
Tomlins et~al. (\citeyear{Tomlins}). We computed the observed test statistic $|S_3|=
|S_3(81)|$ with $N=4$ (Section~\ref{seccrstat}) by counting the number
of genes that were among the top $ 81$ genes in at least $3$ studies
(Table \ref{ex}). As seen in Table \ref{ex}, the set $S_3(81)$
contained 1 ``hit.''
To compute the associated $p$-value, we obtained the expected value of
$|S_3(81)|$ under the null hypothesis as $10\mbox{,}000 P_3^0 (81) = 0.021$
using equation~(\ref{P0}). Then the probability that a Poisson variate
with mean $ 0.021$ will exceed 4 is $0.02$, giving the $p$-value reported
in Table \ref{ex}. Thus, we declare the set $S_3(81)$ to be a
statistically significant set. The single\vadjust{\goodbreak} gene in the set~$S_3(81)$ is
ERG, a gene also found by Tomlins et al. in their study, and this would
be the single gene recommended for further investigation from our
study. At the stringent within-set FDR, we would have confidence at
about the 95\% level that this was not a false positive result.

To gain additional insight, Table \ref{ex} presents $p$-values and FDRs
for recaptured sets over a range of thresholds $ r$ and recapture rates
$n$. Note that only $S_3(81)$ (Table \ref{ex}) is considered a
discovery according to our prespecified analysis strategy; other sets
could be presented as exploratory descriptive results. For completeness
we also examined the four genes corresponding to $|S_2(25)|$, as this
had a highly significant $p$-value and a reasonable 9\% FDR: these are
ERG, ETV1, EST and VGLL3, of which the first two were validated as
participants in a fusion gene by Tomlins et~al. (\citeyear{Tomlins}). Thus, by
setting our FDR to the stringent level of 0.01, we accomplished the
goal of identifying a significant set which contained no false
discoveries, however, we missed one of the truly positive genes. Since
validation of such bioinformatic searches using rtPCR or other
experimental techniques is expected, applying a less stringent a priori
FDR may be a reasonable approach.

To investigate the adequacy of the Poisson approximation, the $p$-values
in Table~\ref{ex} were also verified by direct simulation as in Supplement
Section 2.1 [\citet{Natarajan}]. The $p$-value for the observed $|S_3(81)| =
1$ (Table \ref{ex}) via direct simulation was $0.0188$ compared to the
Poisson approximation $p$-value of $0.0209$. Considering the observed
counts $|S_2(r)|$ in Table~\ref{ex}, the $p$-values derived from the
simulated null distribution of $S_2(r)$ were $0.0775, 0.2001, 0.0602,
0.0004$ and $0.0018$, respectively, for the corresponding thresholds
$r$ of $100, 81, 50, 25$ and $10$. Thus, the Poisson approximation
$p$-values and simulated $p$-values show good concordance.

\section{Discussion} \label{conclusion}
\subsection{Dependence between genes}
\label{seccorr}

In this paper we have assumed independence of the genes within each
study, however, in fact, expression levels may be positively or
negatively correlated between genes. Importantly, our strategy for
study design (i.e., for choice of $r$ and $n$, given in Section \ref
{design}) depends only the mean of the test statistic $|S_n|$, which is
unchanged under arbitrary dependence. [To see this, note that if genes
$g$ and $h$ are correlated, (\ref{ESn}) and (\ref{P0}) still hold.]
However, correlation between genes will induce correlation between the
Bernoulli trials in expression (\ref{distSn}), and, thus, $p$-values
computed under the assumption of independence under the null hypothesis
may no longer be correct. How to adjust for correlation between genes
in the analysis of gene expression studies is an active area of
research [\citet{Efron}, \citet{Benjamini01}, \citet{Sun}]. Here we give some quantitative
guidance in the current setting, using simulation and by considering
theoretical cases of extreme dependence. More detailed analysis will be
the subject of future work.\vadjust{\goodbreak}

First, it is easy to see that negative correlation between genes may be
generally expected to reduce, and positive correlation to increase,
$p$-values as compared to the independent case. This is because
correlation between genes will induce correlation between the Bernoulli
trials $B_g$ in (\ref{distSn}). The variance of the sum will be
correspondingly decreased or increased with the mean remaining
unchanged, rendering the distribution of $|S_n(r)|$ either more or less
concentrated about its mean. It thus is most important to consider the
effect of positive correlation between genes because this will
potentially increase $p$-values and thus type I error, if $p$-values are
computed under the (incorrect) assumption of independence. The
magnitude of the perturbation to $p$-values clearly depends on both the
number of correlated genes and the strength of their correlation, while
the exact perturbation depends on the joint distribution of the
correlated genes. Simulation studies reported in the supplemental
article (Section 1.2) [\citet{Natarajan}] show that moderate correlation
(half of genes with weak correlation or a~few genes with strong
correlation) does not appear to appreciably affect $p$-values. Further
support for these observations is found in the literature on models for
correlated Bernoulli trials, of which [\citet{Yu}, \citet{Gupta}] give relevant examples.

Quantitative insight on the potential magnitude of a perturbation can
be gained by considering the following extreme model: suppose the $T$
genes can be partitioned into modules of size $m$, where two genes
within a module have correlation $\rho>0$ but any two genes in
different modules are independent. In the limiting case with $\rho=1$,
it is easy to see that $|S_n(r)|$ has the same null distribution as the
statistic $|m\widetilde{S}_n(r/m)|$, where the distribution of $|
\widetilde{S}_n|$ is computed using $\widetilde{T}=T/m$ independent
genes, and is thus approximately Poisson with mean $ T P^0_n/m$. It
follows that, under this model, $ |S_n(r)|$ has unchanged mean,
variance inflated by a factor of $m$, and that corrected probabilities
can be computed using the relation
%
\begin{equation}\label{eqcorr}
P(|S_n|>x )=P(| \widetilde{S}_n|>x/m).
\end{equation}
As the postulated within-module correlation decreases from $\rho=1$
toward zero, the correct tail probabilities will smoothly interpolate
from the correction given in (\ref{eqcorr}) to the original values as
computed in Section \ref{secnull} under independence. Thus, given
correlated gene modules of postulated maximum size $m$, relation (\ref
{eqcorr}) might be used to give a conservative ballpark correction to
computed $p$-values.

An example can be computed using the data in Table \ref{ex}, Section
\ref{secexample}. For example, four genes were recaptured by $S_2(25)$
(line 1 of Table \ref{ex}), which had a $p$-value of 0.0006 under the null
hypothesis of independent random ranking of genes. Suppose we wondered
if correlation under the null hypothesis would be sufficient to account
for the observed data. After consideration, suppose we decided there
were several pairs of strongly correlated genes, so that we wanted to
conservatively adjust for many correlated gene modules of size 2, so
$m=2$. As before, $p_0= r/T = 25/10\mbox{,}000$ and $P^0_2 = P(X\geq2)= 3.74
\times10^{-5}$, so that under independence $| {S}_2(25)|$ is
approximately Poisson with mean $ T P^0_n = 0.374$. However, under the
correlated model with $\rho=1$, $| {S}_2(25)|$ is distributed as $| 2
\widetilde{S}_2(25)|$, where $| \widetilde{S}_2(25)|$ is approximately
Poisson with mean $ T P^0_n /2 = 0.187$. The $p$-value adjusted for
correlation would be $P(| \widetilde {S}_n|\geq4/2)= 0.015$. Thus, the
maximum correlation assuming pair-wise modules would be unable to
completely account for the observed data. Several simulation examples
showing the effect of other correlation structures are presented in the
Supplement [\citet{Natarajan}].

\subsection{Conclusions and future directions}

Public repositories of genomic data continue to grow, and
list-intersection approaches similar to those considered here are
likely to become even more common in the future, as several
repositories of curated gene lists have recently been published which
include tools for comparing lists and intersecting lists of top-ranked
genes across multiple similar studies [\citet{Glez-Pena},
\citet{Culhane}]. The primary statistical challenges for analyzing
data from such repositories are controlling the number of false
positive results and maintaining a valid basis for inference when
combining multiple studies [\citet{Benjamini09}].

A well-established method for pooling results across multiple studies
is meta-analysis. This approach is usually conducted gene-by-gene, and
produces a combined $p$-value (or effect-size) for each gene
[\citet{Zaykin}, \citet{Benjamini}, \citet{Garrett-Mayer},
\citet{Pyne}]. However, under this approach it
is possible that a significant gene can be declared based on a few
studies which display large effects, with null effects observed in most
studies, and this can lead to high false positive rates [\citet{Pyne}].
Garrett-Mayer and others [\citet{Garrett-Mayer}] discuss the importance of
first identifying genes that are consistently measured across different
microarray platforms, which is clearly a useful preliminary analysis
for reducing false positives.
There is evidence that rank-based approaches may be more robust and
better guard against false discoveries, while maintaining adequate
power, compared to more traditional methods of meta-analysis [\citet
{Hong}]. Formal or informal rank-based meta analyses for combining
effect sizes across multiple studies have been proposed in the applied
and methodological literature [\citet{Chan}, \citet{Jeffries},
\citet{Deng}, \citet{Miller}].

Our approach compares within-study ranks to a common threshold, and is
an effort to explore the inferential basis of the list-intersection
approach. We provide exact formulas which allow examination of power
and false discovery rates. Our rank-threshold method does not combine
individual per-gene effect sizes, such as ranks, across multiple
studies. Instead, we evaluate the entire set of genes recaptured as
above a rank threshold across multiple studies. Loosely speaking, this
is akin to acceptance sampling procedures, where a ``lot'' [i.e., the
set $S_n(r)$ in our notation] may be deemed ``acceptable'' if the
number of ``defectives'' (i.e., false discoveries) is below some level
(defined by the FDR). However, a salient point in our setup is that
many ``lots'' [$S_n(r)$] may be acceptable, in that they satisfy the
FDR criteria. How to choose among these multiple ``acceptable''
gene-sets is a a major focus of our work. We discuss expected
sensitivity of the gene-sets, and also obtain a $p$-value per recaptured
set, with tight control of the set-wise type I error rate. In this
sense our method is more stringent than an approach which combines
gene-by-gene effect sizes across studies. Our set-based method may also
provide tighter control of gene-level type I error and false discovery
rates, although this is the subject of future research.

In related work [\citet{Pyne}], a pooled $p$-value is calculated together
with a \textit{consensus parameter} defined as the number of studies in
which a feature has to be declared significant before it is considered
significantly validated across studies. Thus, the consensus parameter
plays a similar role as our recapture rate $n$.
\citet{Pyne}
describe results for different values of such consensus parameters but
do not give guidelines on how to choose this parameter. Our work
provides the applied practitioner with $p$-values and expected number of
false positives under various choices for within-study significance
thresholds $r$ and recapture rates~$n$, which could be used to guide
decisions on significant ``gene-sets.''

Another method, the partial conjunction hypothesis test [\citet
{Benjamini}], uses a $p$-value threshold to consider among how many
studies out of $N$ a given gene is found to be significant at a given
level, where each study addresses a different research hypothesis. For
each gene $g$, the set of hypotheses that the gene is null in $n$ or
fewer studies is simultaneously tested, for $1 \le n \le N$. A general
data-driven method for controlling the FDR across all genes is
presented, where the number of false discoveries is defined as the
number of genes which have been called significant in at least one
study in which the gene was truly null. In this setting the studies may
address differing alternative hypotheses, and the focus is on the
situation where a gene can be truly null in some but not other studies,
and where this may differ from gene to gene. Thus, it is of interest to
ascertain for each gene in which studies among all $N$ considered it is
truly significant. This differs from the scenario considered in the
present paper, in which the studies are assumed to each test the same
hypothesis. In the setting of Benjamini et al. power necessarily
declines as the total number of studies $N$ increases
[\citet{Benjamini}]. This is in contrast to our Figure
\ref{figdesign}, in which the expected sensitivity increases with the
total number of studies~$N$. Benjamini et al. have the advantage,
however, of controlling the false discovery rate in a data-driven
manner. By contrast, we allow the user to set the number of studies $n$
that a gene is required to be captured by, and we study how the
expected true positive proportion and false discovery rates are
affected by which alternative is considered to hold. Our interest is in
scenarios where the alternative hypothesis is the same across studies,
as is often the case in genomic studies.

In another approach to the problem, \citet{Lu} develop a
bootstrap methodology for assessing the average frequency with which
significant genes will be rediscovered under independent validation.
This approach is useful at the end of a study when significant
gene-lists have been identified, and no external validation data set is
available. In particular, it can be used to estimate the internal
stability of discovered genes, and also to compare different ranking
procedures applied within the same study. Our focus is on external
validation. We aim to provide a formal statistical method for
evaluating genes that replicate across multiple studies. We discuss how
one might a priori choose within-study significance thresholds
(i.e., $ r$) and cross-study recapture rates (i.e., $n$) to ensure (i)
adequate probability of capturing true positives, and (ii) low false
discovery rate within the recaptured set, when designing a
bioinformatic search across multiple genomic data sets. After this
search is complete, our methods can be applied to obtain $p$-values for
observed recaptured sets, although permutation tests could also be used
to obtain $p$-values, while the bootstrap could be used to obtain
distributions of test statistics.

Our approach has some limitations. We assume that the threshold $ r$
for determining high-ranking genes is the same for all studies, as are
the corresponding probabilities $p_0$ of selecting a null gene (used in
computing $p$-values) and $p_a$ of selecting a differentially expressed
gene (used in computing the expected proportion of true positives for
study design). These assumptions could be relaxed computationally,
although the distributional calculations would lose their simple closed
form solutions. Often when comparing results across studies the
technology used to generate the data will be similar, in which case
requiring similar parameters across studies should not pose a serious
problem. In fact, this assumption is analogous to the homogeneity test
in meta-analysis where only studies with similar design, populations,
and measurement methods are pooled. Further, as with many methods in
common use in the analysis of gene-expression data, our calculations
assume that genes are independent, which is unlikely to be the case in
practice, as discussed in Section \ref{seccorr}. As with any analysis
of gene-expression data using microarrays, RNA-seq, or other
technologies, it is expected that results will be independently
verified using different experimental methods.\vadjust{\goodbreak}

In summary, in this article we describe a simple and rigorous
inferential method for evaluating the consistency of results across
multiple independent studies, using a combined type I error for
discovery of a significant gene set, and an estimated FDR within the
gene set. We show how to choose study parameters to maximize the
expected number of significant genes that will be captured. Future work
will consider related approaches which are based on FDR control. The
framework we describe for selecting a significant set of genes is used
widely by biologists and bioinformaticians [\citet{Tomlins},
\citet{Pyne}, \citet{Benjamini09},
\citet{Glez-Pena}, \citet{Culhane}]. We hope that providing a simple
computational and statistical underpinning for such studies will lead
to more formal use of these methods with corresponding improved control
of type I error rates.

\begin{supplement}[id=suppA]
\stitle{Online supplement: Statistical tests for the intersection of
independent lists of genes}
\slink[doi]{10.1214/11-AOAS510SUPP} 
\slink[url]{http://lib.stat.cmu.edu/aoas/510/supplement.pdf}
\sdatatype{.pdf}
\sdescription{Simulation studies and proofs are in the online
supplement. In Section S1 we show by simulation that the Poisson
approximation to the null distribution of the test statistic gives
reliable $p$-values under a wide range of parameters, both for the
independent case (Section S1.1) and under a range of moderate positive
correlation structures (Section S1.2). We confirm that the Poisson
approximation computed under assumed independence yields conservative
$p$-values under examples of extreme positive correlation, as conjectured
in the text (Section 6.1). In Section S2 we derive the alternative
distribution of the test statistic for some useful special cases, using
combinatorial results \citet{Feller}.}
\end{supplement}

%

\printaddresses

\end{document}